\documentclass[12pt]{article}
\usepackage{latexsym}
\usepackage{amssymb}
\usepackage{float}
\usepackage{amsmath}
\usepackage{psfrag,graphicx,verbatim,array,enumerate}
\usepackage{amsmath,amsfonts,bm,rotating,natbib}
\usepackage{pstricks,pst-node}
\usepackage{epsfig}
\usepackage{pgf}
\usepackage[footnotesize,bf]{caption}

\DeclareMathOperator*{\argmin}{argmin}

\def\balpha{\pmb{\alpha}}
\def\bgamma{\pmb{\gamma}}

\def\by{\pmb{y}}

\begin{document}


\begin{center}

\renewcommand{\thefootnote}{\fnsymbol{footnote}}

{\bf
A Bayesian Variable Selection Approach to Major League Baseball Hitting Metrics\footnote{Blake McShane, Alex Braunstein, and James Piette are doctoral candidates and Shane Jensen is an Assistant Professor, all in the Department of Statistics at the Wharton School of the University of Pennsylvania.  All correspondence on this manuscript should be sent to Blake McShane, \texttt{mcshaneb@wharton.upenn.edu}, 400 Jon M. Huntsman Hall, 3730 Walnut Street, Philadelphia, PA 19104.}
}

\vspace{1.5cm}

\renewcommand{\thefootnote}{\arabic{footnote}}
\setcounter{footnote}{0}

Blakeley B. McShane, Alexander Braunstein, James Piette, and Shane T. Jensen
Department of Statistics \\
The Wharton School \\
University of Pennsylvania \\

\end{center}

\bigskip

\begin{abstract}
Numerous statistics have been proposed for the measure of offensive ability in major league baseball. While some of these measures may offer moderate predictive power in certain situations, it is unclear which simple offensive metrics are the most reliable or consistent.  We address this issue with a Bayesian hierarchical model for variable selection to capture which offensive metrics are most predictive within players across time.  Our sophisticated methodology allows for full estimation of the posterior distributions for our parameters and automatically adjusts for multiple testing, providing a distinct advantage over alternative approaches.  We implement our model on a set of 50 different offensive metrics and discuss our results in the context of comparison to other variable selection techniques.  We find that 33/50 metrics demonstrate signal.  However, these metrics are highly correlated with one another and related to traditional notions of performance (e.g., plate discipline, power, and ability to make contact).
\end{abstract}

\vspace{1cm}

\begin{center}
\today
\end{center}

\newpage

\section{Introduction} \label{introduction}

\begin{quotation}
\noindent \emph{I don't understand. All of a sudden, it's not just BA and Runs Scored, it's OBA. And what is with O-P-S?} - Harold Reynolds
\end{quotation}

The past decade has witnessed a dramatic increase in interest in
baseball statistics, as evidenced by the popularity of the  books
\emph{Moneyball} \cite[]{Lew03} and \emph{Curve Ball}
\cite[]{AlbBen03}.   Beyond recent public attention, the
quantitative analysis of baseball continues to be active area of
sophisticated research (e.g. \cite{Jam08}, \cite{KahGolSil09}).
Traditional statistics such as the batting average (BA) are
constantly being supplemented by more complicated modern metrics,
such as the power hitting measure ISO \cite[]{Puer03} or the
base-running measure SPD \cite[]{James87}. The goal of each measure
remains the same: estimation of the true ability of a player on some
relevant dimension against a background of inherent randomness in
outcomes. This paper will provide a statistical framework for
evaluating the reliability of different offensive metrics where
reliability is defined by consistency or predictive performance.

There has been substantial previous research into measures of offensive performance in baseball.  \cite{Sil03b} investigates the randomness of interseason batting average (BA) and finds significant mean reversion among players with unusually high batting averages in individual seasons. \cite{Stu07} used several players to investigate relationships between infield fly balls, line drives and hits.   \cite{Nul09} uses a sophisticated nested Dirichlet distribution to jointly model fourteen batter
measures and finds that statistical performance is mean reverting. \cite{Bau08} uses algebraic relationships to demonstrate the superiority of on-base percentage (OBP) over batting average (BA).

\cite{Stu07b} considers four defense-independent pitching statistics (DIPS) for individual batters: walk rate, strikeout rate,  home-run rate, and batting average on balls in play (BABIP).   These four measures form a sequence where each event is removed from the denominator of the next event, e.g., player can't strike out if he walks, he cannot hit a home run if he walks or strikes out, etc. \cite{Stu07} finds that the first three measures are quite consistent whereas the fourth measure, BABIP, is quite noisy.   This BABIP measure has been modified in many subsequent works.  \cite{Led09} considers BABIP and groundball outs in the 2007-08 seasons and concludes that handedness and position (as a proxy for
speed) are useful for predicting the two measures. \cite{Bro08} builds on this analysis by finding five factors that are predictive of BABIP and groundball outs: the ratio of pulled groundballs to opposite field groundballs, the percentage of grounders hit to center field, speed (Spd), bunt hits per plate appearance, and the ratio of home runs to fly balls.  \cite{Fai08} analyzes the effects of age on various offensive metrics for hitters.  \cite{Kap06} decomposes several offensive statistics into both player and team level variation. and finds that player-level variation accounts for the large majority of observed variation.

Our own contribution focusses on the following question: which offensive metrics are consistent measures of some
aspect of player ability?   We use a Bayesian hierarchical variable selection model to partition metrics into those with predictive power versus those that are overwhelmed by noise.   \cite{ScoBer06} use a similar variable selection approach to perform large-scale analysis of biological data.  They provide a detailed exploration of the  control of multiple testing that is provided by their Bayesian hierarchical framework, which is an advantage shared by our approach.  We implement our model on 50 offensive metrics using MCMC methods.   We present results for several parameters related to the within-player consistency of these offensive measures.  We compare our posterior inference to an alternative variable selection approach involving the Lasso \cite[]{Tib96}, as well as performing a principal component analysis on our results.  We find a large number of metrics (33/50) demonstrate signal and that there is considerable overlap with the results of the Lasso.  However, these 33 metrics are highly correlated with one another and therefore redundant.  Furthermore, many are related to traditional notions of performance (e.g., plate discipline, speed, power, ability to make contact).

\section{Methodology}\label{methodology}

Our goal is a model that can evaluate offensive metrics on their
ability to predict the future performance of an individual player based
on his past performance.   A good metric is one that provides a
consistent measure for that individual, so that his past
performance is indicative of his future performance.   A poor
metric has little predictive power: one would be just as well served
predicting future performance by the overall league average rather than taking into account past individual performance.    We formalize this principle
with a Bayesian variable selection model for separating out players
that are consistently distinct from the overall population on each
offensive measure.  In addition to providing individual-specific
inferences, our model will also provide a global measure of the
signal in each offensive measure.

Our data comes from the \cite[]{Kap09} database.   We have 50
available offense metrics which are outlined in
Appendix~\ref{metriclist}. Our dataset contains 8,596 player-seasons
from 1,575 unique players spanning the 1974-2008 seasons.  It is
worth noting that the data for 10 of the 50 offensive metrics were
not available before the 2002 season, and so for those metrics we
fit our model on 1,935 player-seasons from 585 unique players\footnote{These metrics are BUH, BUH/H, FB/BIP, GB/BIP, GB/FB, HR/FB, IFFB/FB, IFH, IFH/H, and LD/BIP}.   For
a particular offensive metric, we denote our observed data as
$y_{ij}$ which is the metric value for player $i$ during season $j$.
The observed metric values $y_{ij}$ for each player is modeled as
following a normal distribution with underlying individual mean
$(\mu+\alpha_i)$ and variance $w_{ij} \cdot \sigma^2$,
\begin{eqnarray}
y_{ij}  \sim {\rm Normal} (\mu + \alpha_{i} \, , \, w_{ij} \cdot \sigma^2). \label{likelihood}
\end{eqnarray}
The parameter $\mu$ denotes the overall population mean (i.e., Major League Baseball mean) for the given offensive metric, and each $\alpha_i$ are the player-specific differences from that population mean $\mu$.   The weight term $w_{ij}$ addresses the fact that the variance of a season-level offensive metric for player $i$ in season $j$ is a function of the number of opportunities, and so player-seasons with more opportunities should have a lower variance.  As an example, if the offensive metric being modeled is on-base percentage (OBP), then the natural choice for the weight would the inverse of the number of plate appearances (PA).  The weights used for each offensive metric are given in Appendix~\ref{metriclist}.  With this formulation, the parameter $\sigma^2$ represents the global variance of the offensive metric for player-seasons with an average number of opportunities.    The global parameters $\mu$ and $\sigma^2$ are unknown and are given the following prior distributions,
\begin{eqnarray}
\mu \sim {\rm Normal} (0,K^2) \qquad \sigma^2 \sim {\rm Inverse-Gamma} (\alpha_{0},\beta_{0}) \label{prior1}
\end{eqnarray}
where hyperparameter settings of $K^2 = 10000$, $\alpha_0=.01$, $\beta_0=.01$ were used to make these prior distributions non-informative.    We also need to address our unknown player-specific parameters $\alpha_i$.  We could employ a conventional Bayesian random effects model which would place a Normal prior distribution shared by all $\alpha_i$ parameters.  Instead, we propose a more sophisticated model for the unknown individual $\alpha_i$'s that allows differentiation between players that are consistently different from the population mean from those that are not.

\subsection{Bayesian Variable Selection Model}\label{varselection}

We formulate our sample of players as a mixture of (1) "zeroed" players where $\alpha_i = 0$ versus (2) "non-zeroed" players where $\alpha_i \neq 0$.   We use the binary variable $\gamma_i$ to denote the unknown group membership of each player $i$ ($\gamma_i = 0 \Leftrightarrow \alpha_i = 0$; $\gamma_i = 1 \Leftrightarrow \alpha_i \neq 0$).    We denote by $p_1$ the unknown proportion of players that are in the non-zeroed group ($\gamma_i = 1$) and use the prior distribution $\alpha_i \sim {\rm Normal} (0,\tau^2)$ for them.  For the players in the zeroed group, we  have a point-mass at $\alpha_i = 0$.    The variance parameter $\tau^2$ represents the deviations between individual players that have already been deemed to be different from the overall mean.  When $\tau^2$ is large (particularly in relation to $\sigma^2$), this means that there is can be a potentially wide gulf between
zeroed and non-zeroed players.

\begin{figure}
\begin{center}
\pgfimage[width=3in]{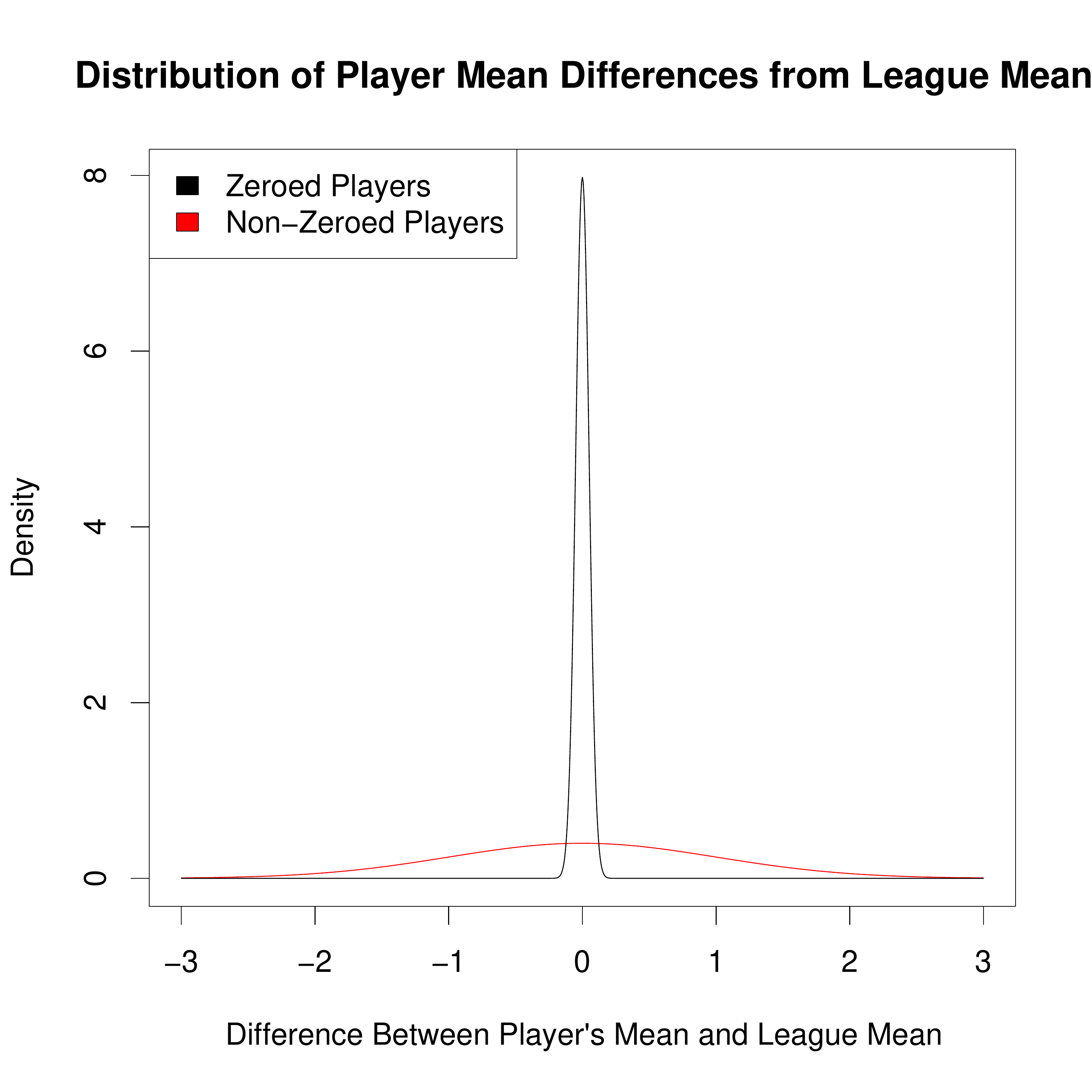}
\caption{Illustration of mixture for player-specific parameters $\alpha_i$.  The black curve approximates a point mass at zero with a normal component that has a very small variance relative to the normal component for the non-zeroed players.}\label{mix}
\end{center}
\end{figure}

\cite{GeoMcC97} demonstrate that using a pure point-mass for a mixture component complicates model implementation.  They suggest approximating the point-mass with a second normal distribution that has a much smaller variance, $v_0 \cdot \tau^2$, where $v_0$ is a hyperparameter set to be quite small.  In our model implementation, we set $v_0 = 0.01$, meaning that the zeroed component has 1/100th of the variance of the non-zeroed component.  Thus, our mixture model on the player-specific parameters is
\begin{eqnarray}
\alpha_i \sim
  \begin{cases}
      {\rm Normal} (0\, , \, \tau^2) & \mbox{if }\gamma_{i}=1 \\
      {\rm Normal} (0\, , \, v_0 \cdot \tau^2) & \mbox{if }\gamma_{i}=0 \\
  \end{cases} \label{alphadist}
\end{eqnarray}
We also illustrate this mixture in Figure~\ref{mix}.
The last two parameters of our model are $\tau^2$ and $p_1$.  We give the following prior distribution to $\tau^2$,
\begin{eqnarray}
\tau^2 \sim  {\rm Inverse-Gamma} (\psi_{0},\delta_{0}).  \label{prior2}
\end{eqnarray}
where the hyperparameters are given values $\psi_0=.01$ and $\delta_0=.01$ in order to be non-informative relative to the likelihood.   \cite{Gel06} cautions that the inverse-Gamma family can actually be surprisingly informative for $\tau^2$ and suggests instead using a uniform prior on $\tau$,
\begin{eqnarray}
p(\tau) \propto 1 \quad \Rightarrow \quad p(\tau^2) \propto 1/\tau \label{prior2.1}
\end{eqnarray}
We implemented this prior as a robustness check and found that our posterior results were nearly identical under these two different prior specifications.

Finally, we allow the mixing proportion parameter $p_1$ to be unknown with prior distribution
\begin{eqnarray}
p_1 \sim {\rm Uniform} (0,1).  \label{prior3}
\end{eqnarray}
As discussed by \cite{ScoBer06}, allowing $p_1$ to be estimated by the data provides an automatic control for multiple comparisons, which is an important advantage of our Bayesian methodology.   Alternative approaches such as standard regression testing of individual means would require an additional adjustment for the large number of tests (1575 players) being performed.


The mixing proportion $p_1$ is also an important model parameter for evaluating the overall reliability of an offensive metric, as it gives the probability that a randomly chosen player shows consistent differences from the population mean.  Therefore, metrics with high signal should have a high $p_1$.  However, a high-signal metric should do more than fit an individual mean to a large fraction of players.  It should also have little uncertainty about which specific players are in the "zeroed" and "non-zeroed" groups.  We can evaluate this aspect of each metric by examining the uncertainty in the posterior distribution of the $\gamma_i$ indicators for each player.  Good metrics should show both a high $p_1$ and a clear separation or consistency in the set of players assigned an individual mean and the set of players assigned the population mean.

\subsection{MCMC Implementation}\label{MCMC}

Let $\by$ be the vector of all player seasons $y_{ij}$ for the given offensive metric $y$.  Similarly, let $\balpha$ and $\bgamma$ denote the vectors of all $\alpha_i$'s and all $\gamma_i$'s respectively.   The use of conjugate prior distributions outlined in Section~\ref{methodology} allows us to implement our model with a Gibbs sampler (\cite{GemGem84}) where each step has a nice analytic form.   Specifically, we iteratively sample from the following conditional distributions of each set of parameters given the current values of the other parameters.

\noindent
{\bf 1. Sampling $\mu$ from $p( \mu | \balpha, \sigma^2, \by)$:}

Letting $i$ index players and $j$ seasons within a player, the conditional distribution for $\mu$ is
\begin{eqnarray*}
\mu | \balpha, \sigma^2, \by \, \sim \, {\rm Normal} \left( \frac{\sum\limits_{i,j} \frac{y_{ij} - \alpha_i}{ w_{ij} \cdot \sigma^2}}{\sum\limits_{i,j} \frac{1}{w_{ij} \cdot \sigma^2} + \frac{1}{K^2}} \, , \, \frac{1}{\sum\limits_{i,j} \frac{1}{w_{ij} \cdot \sigma^2} + \frac{1}{K^2}} \right)
\end{eqnarray*}

\noindent
{\bf 2. Sampling $\balpha$ from $p( \balpha | \mu, \bgamma, \sigma^2, \tau^2, \by)$:}

Again letting $i$ index players and $j$ seasons within a player, the conditional distribution for each $\alpha_i$ is
\begin{eqnarray*}
\alpha_i | \mu, \gamma_i, \sigma^2, \tau^2, \by \, \sim \,
    {\rm Normal} \left( \frac{\sum\limits_{j} \frac{y_{ij} - \mu}{ w_{ij} \cdot \sigma^2}}{\sum\limits_{j} \frac{1}{w_{ij} \cdot \sigma^2} + \frac{1}{\tau_i^2}} \, , \, \frac{1}{\sum\limits_{j} \frac{1}{w_{ij} \cdot \sigma^2} + \frac{1}{\tau_i^2}} \right)
\end{eqnarray*}
where $\tau_i^2 = \tau^2$ if $\gamma_i = 1$ or $\tau_i^2 = v_0 \cdot \tau^2$ if $\gamma_i = 0$.

\noindent
{\bf 3. Sampling $\sigma^2$ from $p( \sigma^2 | \mu, \balpha, \by)$:}

Letting $N$ be the total number of all observed player-seasons, the conditional distribution for $\sigma^2$ is
\begin{eqnarray*}
\sigma^2 | \mu, \balpha, \by \, \sim \,
     {\rm Inv-Gamma} \left( \alpha_0 + \frac{N}{2} \, , \,
               \beta_0 + \sum\limits_{i,j} \frac{(y_{ij} - \alpha_i - \mu)^2}{2 \cdot w_{i,j}}
     \right)
\end{eqnarray*}

\noindent
{\bf 4. Sampling $\tau^2$ from $p( \tau^2 | \balpha)$:}

Letting $m$ be the number of players, the conditional distribution for $\tau^2$ is
\begin{eqnarray*}
\tau^2 | \balpha \, \sim \,
     {\rm Inv-Gamma} \left( \psi_0 + \frac{m}{2} \, , \,
               \delta_0 + \sum\limits_{i} \frac{\alpha_i^2}{2 \cdot v_i}
     \right)
\end{eqnarray*}
where $v_i = 1$ when $\gamma_i = 1$ and $v_i = v_0$ if $\gamma_i = 0$.

\noindent
{\bf 5. Sampling $\bgamma$ from $p( \bgamma | \balpha, \tau^2, p_1) $:}

We sample each $\gamma_i$ as a Bernoulli draw with probability
\begin{eqnarray*}
p(\gamma_i=1 | \alpha_i, \tau^2, p_1) =
\frac{p_1 \cdot \exp\left(-\frac{\alpha_i^2}{2\tau^2}\right)}
{\frac{(1 - p_1)}{\sqrt{v_{0}}}\cdot \exp\left(-\frac{\alpha_i^2}{2 v_{0} \tau^2 }\right) + p_1 \cdot \exp\left(-\frac{\alpha_i^2}{2\tau^2}\right)}
\end{eqnarray*}

\noindent
{\bf 5. Sampling $p_1$ from $p( p_1 | \bgamma)$:}

The mixing proportion $p_1$ has the conditional distribution:
\begin{eqnarray*}
p_1 | \bgamma \, \sim \, {\rm Beta} \left( 1 + \sum\limits_i \gamma_i \, , \, 1 + \sum\limits_i (1-\gamma_i) \right)
\end{eqnarray*}

For each offensive metric, we run our Gibbs sampler for 60,000 iterations and discard the first 10,000 iterations as burn-in.   The remainder of the chain is thinned to retain every 50th iteration in order to eliminate autocorrelation of the sampled values.   We present our results from our estimated posterior distributions in Section~\ref{results}.

\section{Results} \label{results}

We implemented our Bayesian variable selection model on the 50 available offense metrics outlined in Appendix~\ref{metriclist}.   However, we first examined the observed data distribution for each of these metrics to see if our normality assumption (\ref{likelihood}) is reasonable.   The majority of offensive metrics (36/50) have data distributions that are approximately normal.  However, a smaller subset of metrics (14/50) do not have an approximately normal distribution but rather exhibit substantial skewness\footnote{These metrics are 3B, 3B/PA, BUH, BUH/H, CS, CS/OB, HBP, HDP/PA, IBB, IBB/PA, SB, SB/OB, SBPA, and SH}.   Examples are triples (3B) and stolen bases (SB) where the vast majority of players have very small values but there also exists a long right tail consisting of a small number of players with much larger values.   The large proportion of zero values also makes many of these metrics less amenable to transformation.   We proceeded to  to implement our model on all 50 measures, but in the results that follow we will differentiate between those measures that fit the normality assumption versus those that do not.

\subsection{Evaluating Signal in Each Offensive Measure}

As discussed in Section~\ref{varselection}, there are two aspects of our posterior results which are relevant for evaluating the overall signal in an offensive metric.    The first aspect is the proportion of players $p_1$ that have individual means which differ from the population mean.   If a metric has low signal, the population mean has similar predictive power to an individual mean.  We prefer metrics where the individual mean has much more predictive power than the population mean for most players and a good proxy for this characteristic is $p_1$.   Thus, for each metric, we calculate the posterior mean $\widehat p_1$ of the $p_1$ parameter.

In addition to having a large number of players with individual means (large $p_1$), we also want a metric that has high certainty about {\it which} players are consistently different from the overall mean.  Thus, the second aspect of our posterior results that we use to evaluate each metric is the amount of uncertainty in the posterior distributions of the player-specific $\gamma_i$ indicators.  Specifically, a good metric should have posterior estimates where ${\rm P} (\gamma_i = 1) \approx 0$ or ${\rm P} (\gamma_i = 1) \approx 1$ for many players $i$.    A good global summary of this aspect of our model is the negative entropy $-H$,
\begin{equation*}
-H = \frac{1}{m} \sum_{i=1}^{m} [\widehat{\gamma_i} \log(\widehat{\gamma_i}) + (1 - \widehat{\gamma_i}) \log(1 - \widehat{\gamma_i})]
\end{equation*}
where $\widehat{\gamma_i}$ is the fraction of posterior samples where $\gamma_i = 1$ for player $i$.   The negative entropy is maximized (at $-H = 0$) when each $\widehat{\gamma_i}$ is either 0 or 1, which is the ideal situation for a consistent offensive metric.   Correspondingly, the negative entropy is minimized when each $\widehat{\gamma_i} \approx 0.5$ which suggests large uncertainty about each player.

\begin{figure}
\begin{center}
\pgfimage[width=3in]{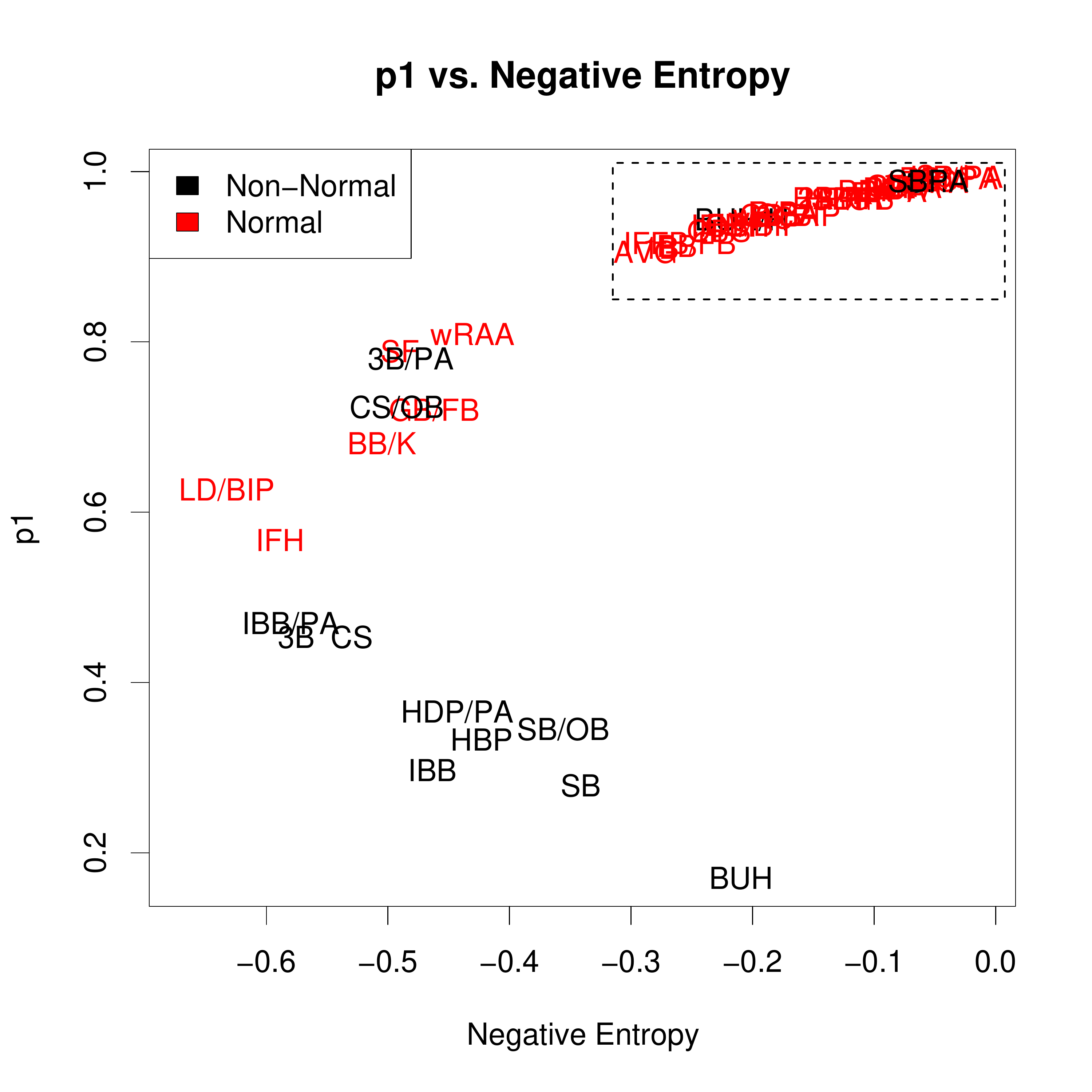}\pgfimage[width=3in]{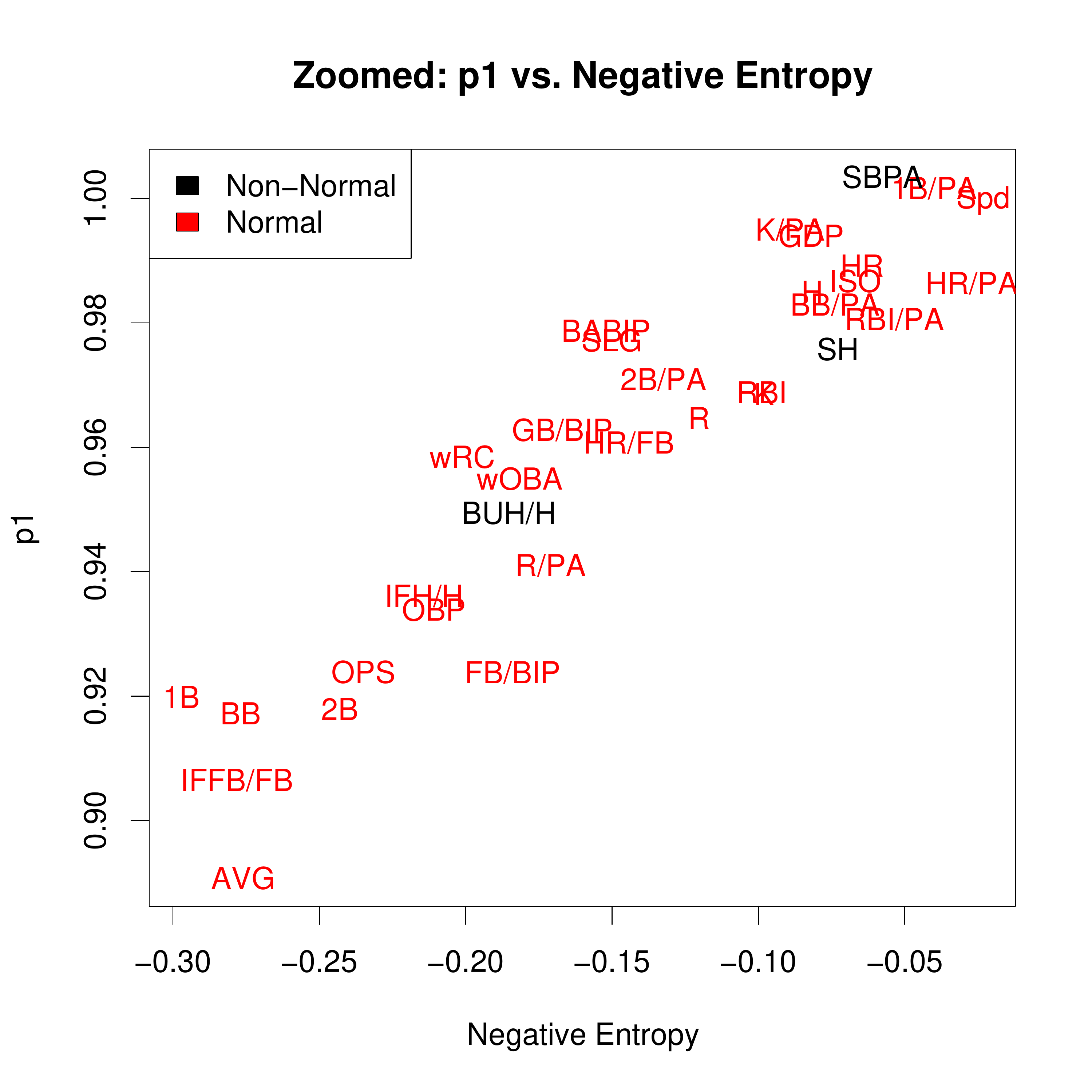}\\
\caption{Plots of $\widehat p_1$ (y-axis) against the negative entropy.  On the right hand plot, we zoom in on the subset of the data enclosed by the dotted rectangle in the upper right portion of the left plot and jitter the points for visibility. }\label{p1_entropy}
\end{center}
\end{figure}

On the left-hand side of Figure~\ref{p1_entropy}, we plot $\widehat p_1$ against the negative entropy $-H$ for our 50 offensive measures.  Metrics colored in red were the majority that were reasonably approximated by a normal distribution, whereas metrics colored in black were not.  We see three groupings of metrics in the left-hand side of Figure~\ref{p1_entropy}.   We see a cluster of black (non-normal) metrics with low values of $\widehat p_1$ which look to be poor metrics by our evaluation but are also a poor fit to our model assumptions.  We also see a cluster of mostly red (normal) metrics that show intermediary  $\widehat p_1$ values but low values of the negative entropy.   These red metrics meet our model assumptions but the results indicate that they are poor metrics in terms of within-player consistency.   The most interesting cluster are the predominantly red metrics that have both large $\widehat p_1$ and large values of negative entropy.   These are the metrics that seem to perform well in terms of our evaluation criteria, and we zoom in on this group on the right-hand side of Figure~\ref{p1_entropy}.    Within this group of 33 high signal metrics, we also see a fairly strong relationship between  $\widehat p_1$ and negative entropy.

Several of the best metrics suggested by our results are K/PA, Spd, ISO, BB/PA, and GB/BIP.  The set of these best metrics spans several different aspects of hitting.  K/PA and BB/PA are all related to plate discipline, Spd represents speed, ISO measures hitting power, and GB/BIP captures the tendency to hit ground balls.  These findings are partly supported by \cite{Stu07b} who finds that K rate and BB rate are very consistent.   However,   \cite{Stu07b} also found that BABIP was a low signal metric, whereas our evaluation places BABIP among the high signal metrics.     In addition to isolating a subset of good metrics, our evaluation based on both $\widehat p_1$ and negative entropy provides a continuum in the right-hand plot of Figure~\ref{p1_entropy} upon which we can further examine these good metrics.

\subsection{Examining Individual Players}

Four metrics found by our model to be high signal were ISO, BB rate, Spd and K rate.  Each of these metrics measures a different aspect of offensive ability: ISO relates to hitting power, BB rate relates to plate discipline, Spd relates to speed, and K rate relates to the ability to make contact.    We further explore our results by focusing on the top individual players for each of these measures, as estimated by our model.   In Table~\ref{topplayers}, we show the top five players in terms of their estimated individual means ($\mu+\alpha_i$) for each of these metrics.

\begin{table}[t]
\begin{center}
\begin{tabular}{|l|cc||l|cc|}
\hline
\multicolumn{3}{|c||}{{\bf ISO - Isolated Power}} & \multicolumn{3}{|c|}{{\bf BB (walk) rate}}  \\
\hline
{\bf Player} & \multicolumn{2}{|c||}{{\bf Mean $(\mu + \alpha_i)$}}  & {\bf Player} & \multicolumn{2}{|c|}{{\bf Mean $(\mu + \alpha_i)$}}   \\
& {\bf Estimate} & {\bf SD} &  & {\bf Estimate} & {\bf SD} \\
\hline
Mark McGwire        & 0.320 & 0.010 & Barry Bonds   & 0.204 & 0.004 \\
Barry Bonds     & 0.304 & 0.008 & Gene Tenace   & 0.186 & 0.007\\
Ryan Howard     & 0.293 & 0.016 & Jimmy Wynn    & 0.183 & 0.010 \\
Jim Thome       & 0.287 & 0.009 & Ken Phelps    & 0.176 & 0.011 \\
Albert Pujols   & 0.281 & 0.011 & Jack Cust & 0.176 & 0.012 \\
\hline
\multicolumn{3}{|c||}{Population Mean $\hat{\mu} = 0.142$} & \multicolumn{3}{|c|}{Population Mean $\hat{\mu} = 0.087$} \\
\hline
\hline
\multicolumn{3}{|c||}{{\bf Spd - Speed}} & \multicolumn{3}{|c|}{{\bf K (strikeout) rate}}  \\
\hline
{\bf Player} & \multicolumn{2}{|c||}{{\bf Mean $(\mu + \alpha_i)$}}  & {\bf Player} & \multicolumn{2}{|c|}{{\bf Mean $(\mu + \alpha_i)$}}   \\
& {\bf Estimate} & {\bf SD} &  & {\bf Estimate} & {\bf SD} \\
\hline
Vince Coleman       & 8.55  & 0.30  & Jack Cust         & 0.388 & 0.018 \\
Jose Reyes          & 8.22  & 0.40  & Russell Branyan   & 0.376 & 0.021\\
Carl  Crawford      & 8.14  & 0.36  & Melvin Nieves         & 0.371 & 0.020 \\
Willie Wilson        & 8.13 & 0.25   & Rob Deer & 0.351 & 0.010 \\
Omar Moreno          & 7.89 & 0.31  & Mark Reynolds & 0.347 & 0.018 \\
\hline
\multicolumn{3}{|c||}{Population Mean $\hat{\mu} = 4.11$} & \multicolumn{3}{|c|}{Population Mean $\hat{\mu} = 0.166$} \\
\hline
\end{tabular}
\end{center}
\caption{Top players for four high signal metrics.   For each player, we provide the posterior estimate and posterior standard deviation for their individual mean ($\mu + \alpha_i$).  The estimated $\widehat{\gamma_i}$ was equal to 1.00 for each of these cases.  The posterior estimate of the population mean $\mu$ is also provided for comparison.}\label{topplayers}
\end{table}

For the isolated power (ISO) metric, each of the top five players are well-known hitters that have led the league in home runs at least once during their careers.   Even more striking is the magnitude of their estimated individual means ($\mu + \alpha_i$), which are more than double the population mean $\mu = 0.142$.  Barry Bonds appears in the top 5 baseball players for both ISO and BB rate, and more generally, there is fairly strong correspondence between these two metrics beyond the results of Table~\ref{topplayers}.   This finding suggests that there is correlation between the skills that determine a batters plate discipline and the skills that lead to hitting for power.  Other well-known players ranking high on BB rate (but outside of the top 5) are Jim Thome, Mark McGwire, Frank Thomas, and Adam Dunn. Barry Bonds does stand out dramatically with a walk rate that is almost 2\% higher than the next highest player, the equivalent of 12-13 extra walks per season.   This difference seems especially substantial when taking into account the small standard deviation (0.4\%) in Bonds' estimated mean.

Jack Cust appears in the top 5 baseball players for both BB rate and K rate, which is especially interesting since having high BB rate is beneficial whereas having a high K rate is detrimental.  However, it is not particularly surprising, since players with good plate discipline will frequently be in high count situations that can also lead to strike outs.   Cust is especially well-known for having a ``three-outcome" (i.e. walk, strikeout or home run) approach.  Moving beyond the top 5 players, other power hitters such as  Ryan Howard, Adam Dunn, and Jim Thome also exhibit high K rates.   The top players on Bill James' speed metric Spd are a much different set of players than the previous three metrics.  The highest estimated individual mean is held by former Rookie of the Year Vince Coleman, who led the National League in stolen bases from 1985 to 1990.

A general theme of all four metrics examined in Table~\ref{topplayers} is that there is consistency {\it within players}, as indicated by the relatively small standard deviations, but clear evidence of substantial heterogeneity {\it between players} since the top players are estimated to have such a large deviation from the population mean.   These two factors are an
ideal combination for a high signal offensive metric.

\section{External Validation and Principal Components}

In the next subsection, we compare our results to an alternative variable selection approach based upon the Lasso.   We then explore the correlation between offensive metrics with a principal component analysis.

\subsection{Comparison to the Lasso}

The Lasso (\cite{Tib96}) is a penalized least squares regression that uses an $L_1$ penalty on the estimated regression coefficients,
\begin{eqnarray}\label{eq_lasso}
\hat{\beta}_{Lasso} = \argmin_{\hat \beta} \left[ \sum_{i,j} (y_{ij} - X_i \hat \beta)^2 + \lambda \sum_i | \hat \beta_i | \right],
\lambda \geq 0
\end{eqnarray}
The Lasso implementation enforces sparsity on the covariate space by forcing some coefficients to zero, and can therefore be used for variable selection.  A more intuitive reformulation of the Lasso is as a  a minimization of $ \sum_{i,j} (y_{ij} - X_i \hat \beta)^2 $ subject to $ \frac{ \sum_i | \hat \beta_i |}{ \sum_i |\hat \beta_i^{OLS} |} \leq f $, where $\hat \beta_i^{OLS}$ is the coefficient from variable $i$ in the ordinary least squares solution.  The free parameter $f$ is known as the Lasso "fraction" and corresponds to $\lambda$ in Equation \ref{eq_lasso}; $f$ ranges between zero (corresponding to fitting only an overall mean or  $\lambda = \infty$ in Equation \ref{eq_lasso}) and one (corresponding to the ordinary least squares regression solution or  $\lambda = 0$ in Equation \ref{eq_lasso}). 

We apply the Lasso to our problem by centering each offensive metric and then fitting the regression model consisting only of indicators for each player.   Each component of the $\hat \beta$ vector corresponds to the individual mean of given player, and we are interested in which of these individual means are fitted to be different from zero. To select a value of the free parameter $f$, we implemented multiple 5-fold cross validation (CV) by randomly subdividing all player-seasons into 5 groups and repeating ten times. We fit the Lasso on 4 groups and predict out-of -sample on the remaining player-seasons. This analysis is repeated for a fine grid of possible $f$ values ranging between 0 and 1, and we selected the $f$ with the lowest cross-validated average RMSE.  We then fit the Lasso model using this value of $f$.

\begin{figure}
\begin{center}
\pgfimage[width=3in]{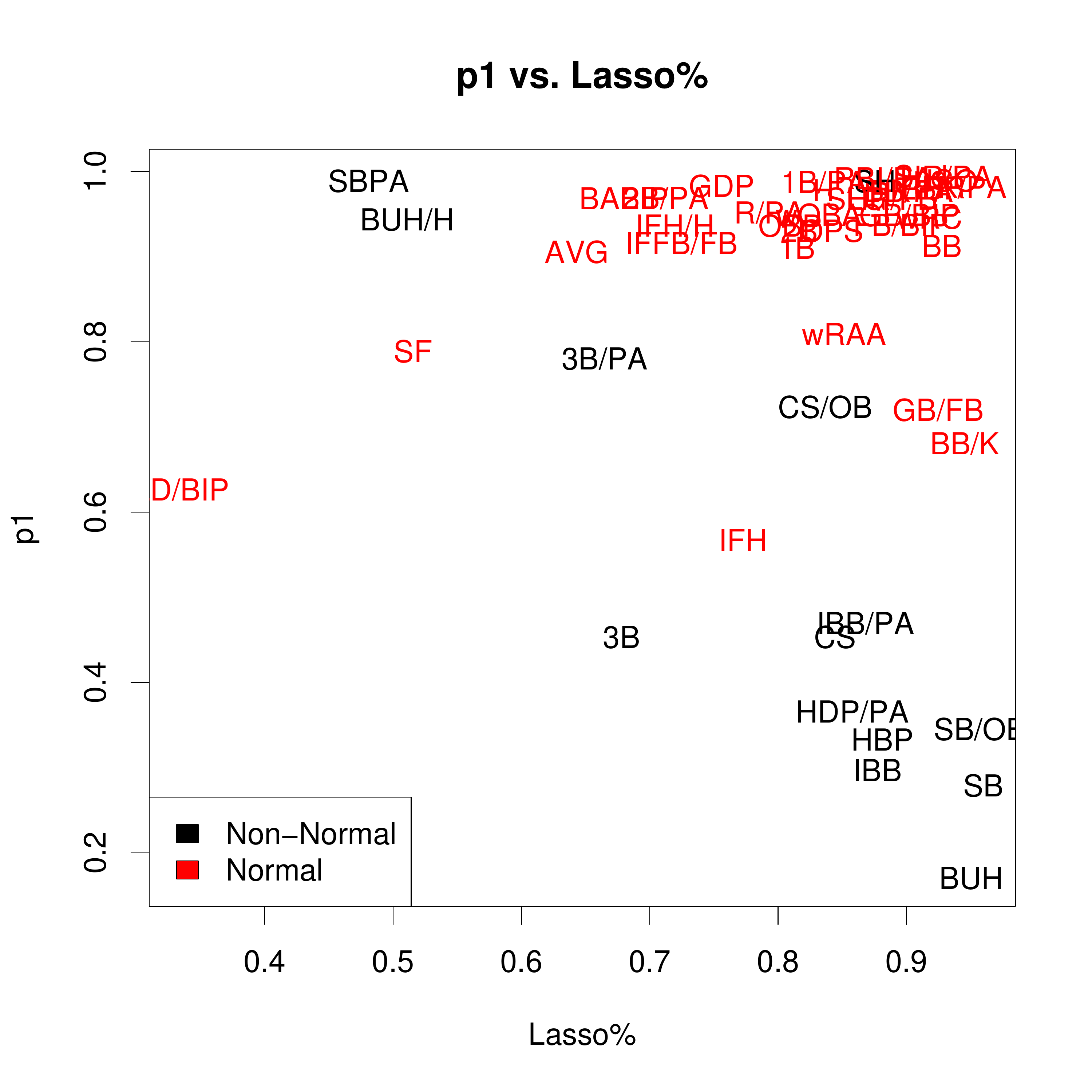}\pgfimage[width=3in]{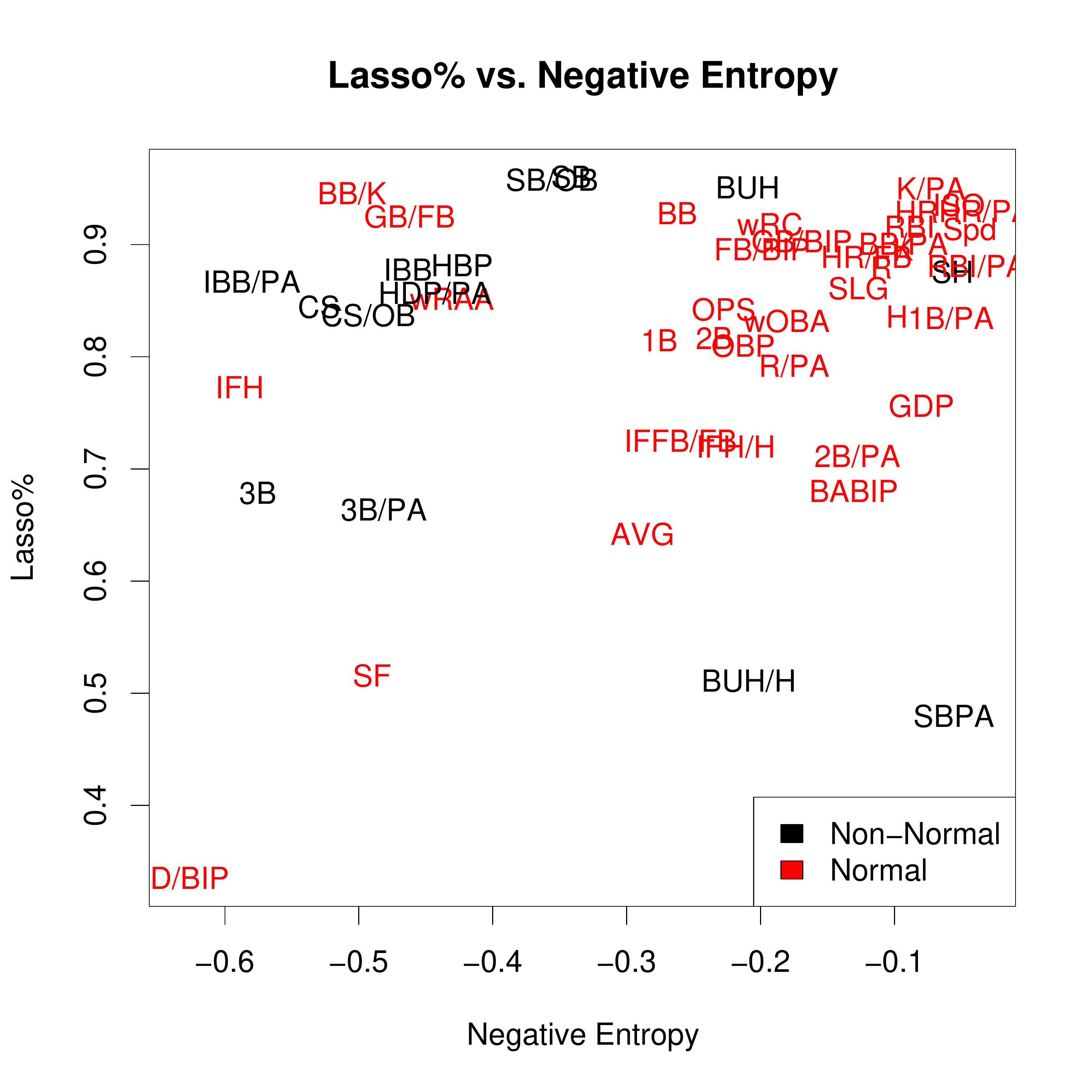}\\
\caption{Left: Plot of $\widehat p_1$ (y-axis) against the percentage of players with non-zero means selected by the Lasso. Right: Plot of percentage of players with non-zero means selected by the Lasso (y-axis) versus negative entropy }\label{lassofig}
\end{center}
\end{figure}

The outcome of interest from this Lasso regression is Lasso\%, the percentage of players that are fitted with non-zero coefficients by the Lasso implementation.   This measure represents a global measure of signal for each metric, and thus serves as an alternative to our model-based measures of $\widehat p_1$ and the negative entropy.   We compare our model-based measures to the Lasso\% measure in Figure~\ref{lassofig}.    In the right plot of Figure~\ref{lassofig}, we see no real structure to the relationship between the negative entropy and the Lasso\%.  This is not surprising considering that the negative entropy is a measure of the variability in our own model-based results, which is not an equivalent measure to the Lasso\%.

More interesting is the comparison of $\widehat p_1$ and Lasso\% as these two measures are more intimately related.
In the left plot of Figure~\ref{lassofig}, we see agreement between Lasso\% and $\widehat p_1$ for many measures, especially the red measures that fit the normal model.   These high signal measures with large value of  $\widehat p_1$ also tend to have a large percentage of non-zero coefficients.   The main difference between the two methods is with the black metrics that have skewed (non-normal) data distributions.   These measures tend to have a high Lasso\% but a low $\widehat p_1$, meaning that a Lasso-based analysis would attribute much more signal to these metrics than our mixture model-based analysis.      Neither our model nor the Lasso is meant for the highly skewed data of these black metrics.  The fact that the results from our model is more cautious about these metrics than the Lasso results suggests an advantage to our approach.

\subsection{Principal Components Analysis}

Among our metrics inferred to have high signal (right-hand plot of Figure~\ref{p1_entropy}), we see a broad and continuous spectrum of performance on both $\widehat{p_1}$ and negative entropy.   Ideally, there would be a more stark divide in the performance of these metrics, allowing us to focus on only a small subset of metrics as a complete summary of offensive performance.   However, this is a difficult task in large part because of the high correlation between many of these metrics.  As an obvious example, OPS is a linear combination of OBP and SLG.  We performed a more systematic assessment of the correlation between metrics using a Principal Components Analysis.   PCA projects the data onto an orthogonal space such that each orthogonal component describes a decreasing amount of variance.

\begin{figure}[t]
\begin{center}
\pgfimage[width=\textwidth]{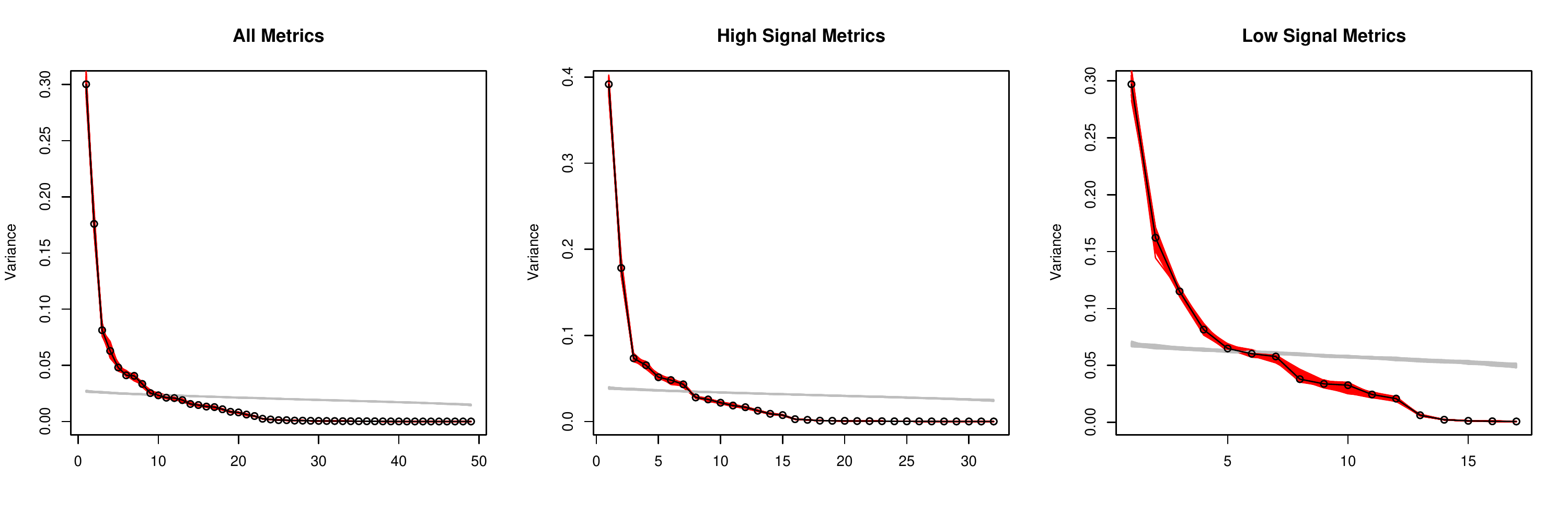}\\
\caption{Plots of the variance explained by each principal component for all metrics (left), for the high signal metrics (middle), and for the remaining metrics (right). For each plot we create a grey null band by randomly permuting the values within each column to demonstrate the strong significance of our results. In addition, we demonstrate the variability in our own principal components by creating bootstrap samples of player seasons and calculating the variance of the bootstrap principal components in red.}\label{pca}
\end{center}
\end{figure}

Note that one of our 50 metrics, SBPA, was not included in this analysis due to a high number of player-seasons which had a denominator (SB+CS) equal to zero.  The results from our principal components analysis on the remaining 49 metrics are shown in Figure~\ref{pca}.    We see that among the 49 metrics represented in the left-hand plot of Figure~\ref{pca}, only about eight principal components have variance exceeding the null bands, which suggests that there are only about eight unique (orthogonal) metrics among the entire set of metrics.

As a further consideration, we computed the principal components of 32 of the 33 high signal metrics\footnote{Again, excluding SBPA.} as well as the principal components of the remainder of the metrics.  If our hypothesis that much of the signal in the data comes from eight significant components,  then we should expect to see about eight signiÞcant principal components for the first set of data with a shape similar to that in the left panel of Figure~\ref{pca}; on the other hand, for the remaining metrics, we expect to see fewer significant principal components as well as a curve which is less steep. As the middle panel of Figure~\ref{pca} shows, there are indeed about six or seven significant principal components in the set of 32 metrics. This means much of the signal in all 50 metrics is contained in the 32 (and, in fact, that there are only really about six or seven truly different ones among those 32). Furthermore, the 17 noisy metrics contain substantially less signal, as shown by the right panel of Figure~\ref{pca}.

\section{Discussion}

We have introduced a hierarchical Bayesian variable selection model, which allows us to determine the ability of metrics for hitting performance to provide sound predictions across time and players.  Our model does not require adjustment for multiple testing across players and imposes shrinkage of player-specific parameters towards the population mean.   For 50 different offensive metrics, the full posterior distributions of our model parameters are estimated with a Gibbs sampling implementation.   We evaluate each of these metrics with several proxies for reliability or consistency, such as the proportion of players found to differ over time from the population mean as well as the posterior uncertainty (in terms of negative entropy) of this deviation from the population mean in individual players.

We find clear separation between 17 metrics with essentially no signal and 33 metrics that range along a continuum of having lesser to greater signal.   The existence of this continuum is largely a function of the high correlation between each of these metrics.   Our principal components analysis suggests that only a small subset of metrics are substantively different from one another.

A direction of future research would be the creation of a reduced set of consistent metrics that were less highly dependent but still directly interpretable.   We also plan to apply our methodology to multiple measures of pitching performance, which we anticipate will be less highly correlated with each other.   Our sophisticated hierarchical model could be extended further to share information between metrics instead of the separate metric-by-metric analysis that we have performed. The relatively high correlation between some metrics could be used to cluster metrics together and reduce dimensionality.  This approach would have the advantage of pooling across related measures and increasing effective degrees of freedom.

\bibliography{references}

\begin{appendix}

\section{Offensive Measures}\label{metriclist}

Our 50 offensive measures are subdivided below into categories for ease of presentation.   Several terms that are not defined in the tables themselves are AB (at bats), BIP (balls in play), OB (total number of times on base), PA (plate appearances), and PA$^\star$ (plate appearances minus sacrifice hits).

{\bf 1. Simple hitting totals and rates}\\
{\footnotesize
\begin{tabular}{|lll||lll|}
\hline
{\bf Metric} $y_{ij}$ & {\bf Weight} $w_{ij}$& {\bf Description} & {\bf Metric} $y_{ij}$ & {\bf Weight} $w_{ij}$& {\bf Description} \\
\hline
1B  & PA     & singles & 1B/PA   & PA & single rate \\
2B  & PA     & doubles & 2B/PA   & PA & double rate \\
3B  & PA     & triples & 3B/PA   & PA & triple rate\\
HR  & PA     & home runs & HR/PA     & PA & home run rate\\
R   & PA     & runs & R/PA   & PA & run rate\\
RBI     & PA     & runs batted in & RBI/PA & PA & runs batted in rate\\
BB  & PA     & base on balls (walk) & BB/PA & PA & walk rate\\
IBB     & PA     & intentional walk & IBB/PA & PA & intentional walk rate\\
K  & PA     & strike outs & K/PA & PA & strike out rate \\
HBP     & PA     & hit by pitch  & HBP/PA & PA & hit by pitch rate \\
BUH        &H    & bunt hits  & BUH/H & H & bunt hit proportion \\
H   & PA     & hits  & GDP     & PA     & ground into double play \\
SF  & PA     & sacrifice fly  & SH  & PA     & sacrifice hit  \\
\hline
\end{tabular}
}

{\bf 2. More complicated hitting totals and rates}\\
{\footnotesize
\begin{tabular}{|lll|}
\hline
{\bf Metric} $y_{ij}$ & {\bf Weight} $w_{ij}$& {\bf Description}  \\
\hline
OBP      & PA$^\star$  & on base percentage (OB/PA$^\star$)\\
AVG     & AB     & batting average (H/AB)\\
SLG     & AB     & slugging percentage\\
OPS     & AB $\times$ PA$^\star$   & OPB + SLG\\
ISO     & AB     & isolated power (SLG-AVG) \\
BB/K & PA    & walk to strikeout ratio \\
HR/FB  &PA   & home run to fly ball ratio\\
GB/FB &BIP   & ground ball to fly ball ratio\\
BABIP & BIP & batting average for balls in play\\
LD/BIP &BIP   & line drive rate\\
GB/BIP  &BIP   & ground ball rate\\
FB/BIP  &BIP   & fly ball rate\\
IFFB/FB &FB   & infield fly ball proportion\\
IFH        &GB   & in field hit\\
IFH/H  &GB   & in field hit proportion\\
wOBA &PA$^\star$  & weighted on base average\\
wRC    &PA  & runs created based on wOBA\\
wRAA &PA     & runs above average based on wOBA\\
\hline
\end{tabular}
}

{\bf 3. Baserunning totals and rates}\\
{\footnotesize
\begin{tabular}{|lll||lll|}
\hline
{\bf Metric} $y_{ij}$ & {\bf Weight} $w_{ij}$& {\bf Description} & {\bf Metric} $y_{ij}$ & {\bf Weight} $w_{ij}$& {\bf Description} \\
\hline
SB  & OB & stolen bases & SB/OB  & OB & stolen base rate \\
CS  & OB & caught stealing & CS/OB  & OB & caught stealing rate \\
SBPA  & SB + CS  & stolen bases per attempt  & Spd     & PA     & Bill James' speed metric \\
 & & \phantom{yo} i.e., SB/(SB+CS) & & & \\
\hline
\end{tabular}
}
\end{appendix}

\end{document}